\documentclass[pra,aps,twocolumn]{revtex4}
\usepackage[12pt]{xy}
\usepackage{graphicx}
\usepackage{colordvi}
\usepackage{amsmath}
\usepackage{color}

\begin{document}

\title{Violation of Bell's inequality using classical measurements
and non-linear local operations}

\author{Magdalena Stobi\'nska,$^1$ Hyunseok Jeong,$^2$ and Timothy C. Ralph$^2$}

 \affiliation{
 $^1$Instytut Fizyki Teoretycznej, 
 Uniwersytet Warszawski, Warszawa 00--681, Poland
\\
$^2$Centre for Quantum Computer
  Technology, Department of Physics, University of Queensland,
  St Lucia, Qld 4072, Australia} 
  
\date{\today}

\begin{abstract}
We find that Bell's inequality can be significantly violated (up to
Tsirelson's bound) with two-mode entangled coherent states using only
homodyne measurements. This requires Kerr nonlinear interactions for
local operations on the entangled coherent states.  Our example is a
demonstration of Bell-inequality violations using {\it classical}
measurements. We conclude that entangled coherent states with coherent
amplitudes as small as $0.842$ are sufficient to produce such
violations.
\end{abstract}


\maketitle

\section{Introduction}

Quantum entanglement is one of the most distinguishing properties of
quantum theory.  It is well known that some entangled states violate
Bell's famous inequality which is imposed by any local-realistic
theory \cite{Bell}.  The coherent states with large amplitudes are
known as most classical among all pure states \cite{coh}, and two
well-separated coherent states in the phase space can be considered
{\it classically} (or macroscopically) distinguishable, i.e. they can
be efficiently discriminated by homodyne detection in quantum optics
without detecting individual quanta.  In this sense, an entangled
coherent state (ECS) can be regarded as an interesting example of
entanglement between classically distinguishable states
\cite{Sanders}.  The ECSs in free-traveling fields have been studied
as useful resources for quantum information processing
\cite{ccr,Enk,JKL01,Wang,BaAnTeleport,JK,Ralph,
  JKpuri,Clausen,Glancy}.  A~single-mode superposition of coherent
states (SCS) can be simply converted to an ECS at a balanced beam splitter.
Recently, experimentally feasible schemes have been suggested to
generate the SCS and the ECSs in
free-traveling fields \cite{g-1,g-15,MP}. Recent experimental progress
shows that the generation of the ECSs is now within reach of current
technology \cite{g-2}.

It was found that violations of Bell's inequality for the ECSs can be
demonstrated using photon detection, i.e. either photon counting
measurements or photon on/off measurements \cite{Wilson,jeongsonkim}.
However, photon detection cannot be considered a classical measurement
as it detects individual photons.  In order to demonstrate
Bell-inequality violations for the ECS as entanglement between
classically distinguishable states, one needs to use measurements
which have more classical nature such as homodyne detection.  It is
also worth noting that homodyne detection can be performed with high
efficiency using current technology compared to photon detection.
There exist proposals for Bell-inequality tests with some continuous
variable states using homodyne detection, but the required states tend
to be quite exotic \cite{hb}.

In this paper we find that Clauser, Horne, Shimony and Holt (CHSH)'s
version \cite{CHSH} of the Bell inequality can be violated up to
Tsirelson's bound $2\sqrt{2}$ \cite{C80} with an ECS using homodyne
detection.  Required local operations may be realized using Kerr
nonlinearities and simple linear optics elements.  An
interesting question answered by our investigation is: how large must
the amplitude of the ECS be in order to violate
Bell's inequality with respect to classical measurements?

This paper is organized as follows. In Sec.~II we study violations of
the Bell-CHSH inequality for ECSs using homodyne detection and
idealized local operations.  This introduces the scheme in a
straightforward way and illustrates the limits introduced specifically
by the homodyne measurement. We then explain in Sec.~III how to
implement the local operations using Kerr
nonlinearities, beam splitters and phase shifters, and
derive new limits to the Bell violation in the presence of physically
realizable local operations.  We conclude with final remarks in
Sec.~IV.

\section{Bell inequality test for~an~ECS using homodyne detection} 

We introduce four ECSs
\begin{eqnarray}
\label{bell1}
  |\Phi_\pm\rangle&=&N_\pm(|\alpha\rangle|\alpha\rangle \pm
  |-\alpha\rangle|-\alpha\rangle), 
\\
\label{bell2}
  |\Psi_\pm\rangle&=&N_{\pm}(|\alpha\rangle|-\alpha\rangle \pm
  |-\alpha\rangle|\alpha\rangle),
\end{eqnarray}
where $N_\pm=\{2(1 \pm e^{-4|\alpha|^2})\}^{-1/2}$ and
$|\alpha\rangle$ is a coherent state with amplitude $\alpha$.
We also define a local operation $\hat R(\varphi)$ as
\begin{equation}
\begin{aligned}
&\hat R(\varphi)|\alpha\rangle = \cos \varphi  |\alpha\rangle +
\sin \varphi |-\alpha\rangle, 
\\
&\hat R(\varphi)|-\alpha\rangle = \sin \varphi |\alpha\rangle
 - \cos \varphi |-\alpha\rangle,
 \label{e:R}
\end{aligned}
\end{equation}
which is nonunitary due to the non-orthogonality of $|\alpha\rangle$
and $|-\alpha\rangle$.  However, $\hat R(\varphi)$ becomes
approximately unitary when the overlap between the two coherent
states, $\langle \alpha|-\alpha \rangle = e^{-2|\alpha|^2}$,
approaches zero.  It should be noted that this overlap goes rapidly to
zero as $\alpha$ increases.

\begin{figure}[h!]
\begin{center}
\scalebox{0.45}{\includegraphics{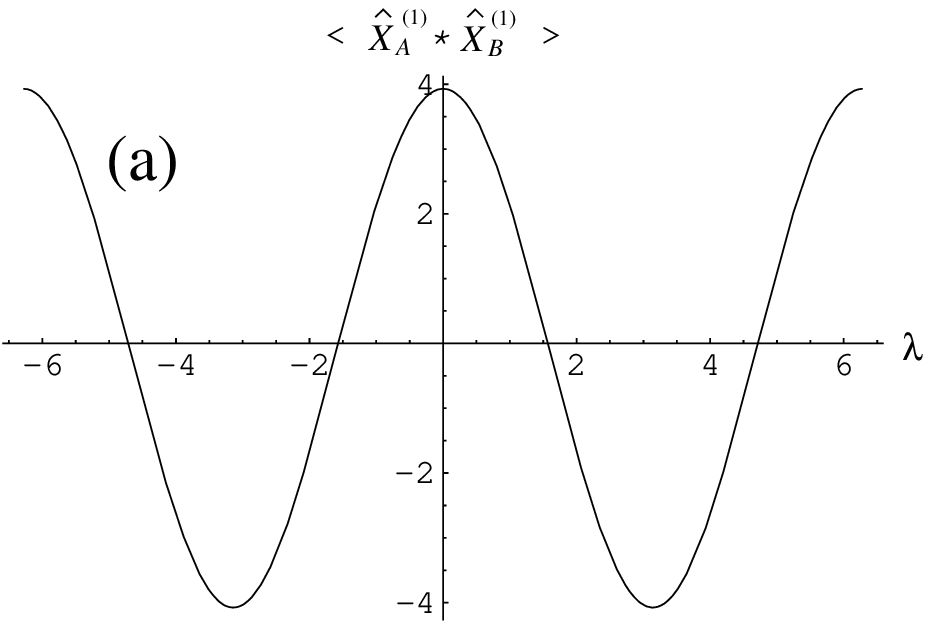}
{\includegraphics{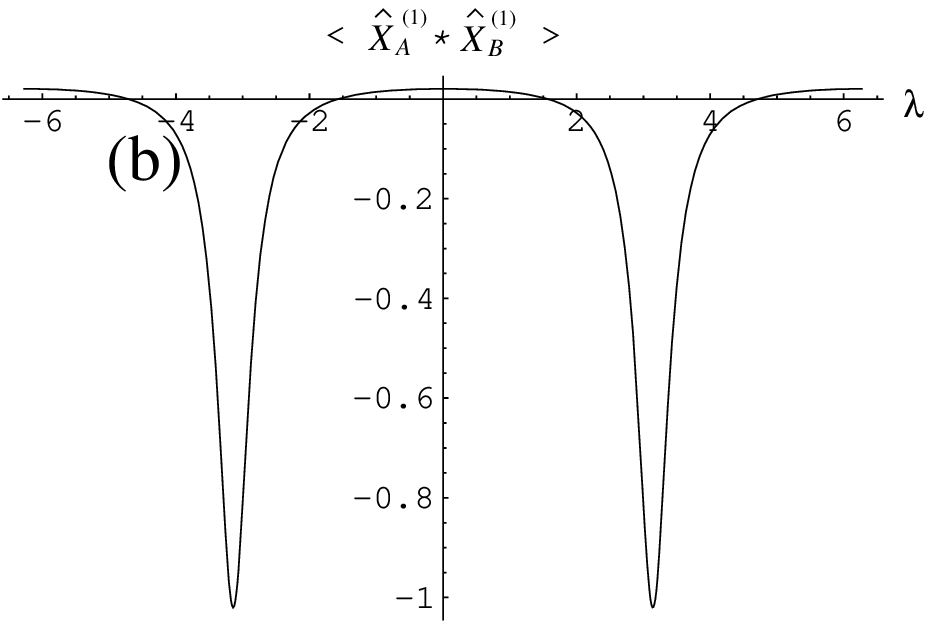}}}
\caption{The amplitude quadrature correlations, $\langle \hat
  X_{A}^{(1)} \hat X_{B}^{(1)}\rangle$, for (a) $\alpha = 1$ and (b)
  $\alpha =0.1$ against $\lambda$, where $\lambda = 2(\phi -
  \theta)$.}
\label{quadratue_1}
\end{center}
\end{figure}
\begin{figure}[h!]
\begin{center}
\scalebox{0.45}{\includegraphics{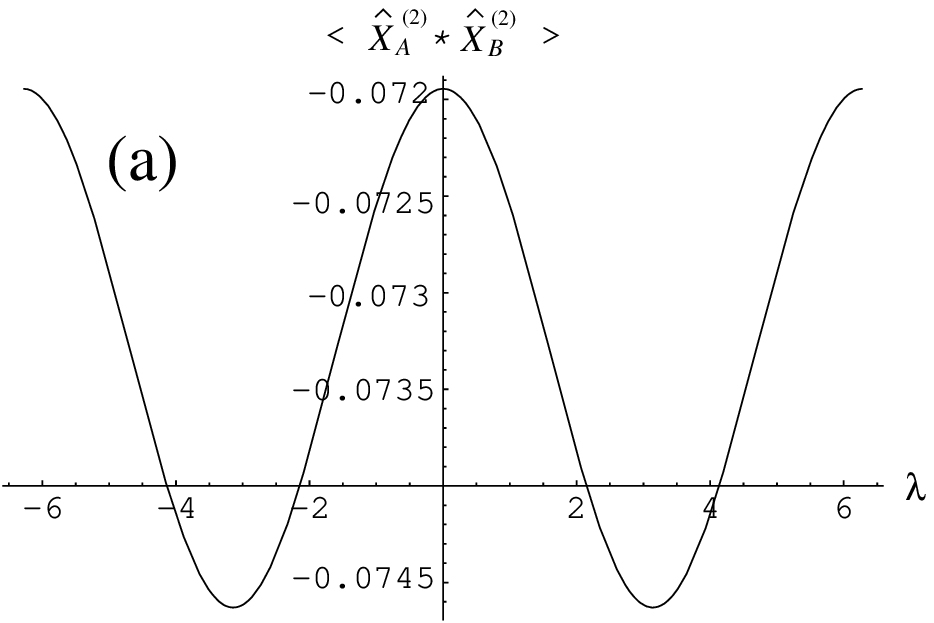}
\includegraphics{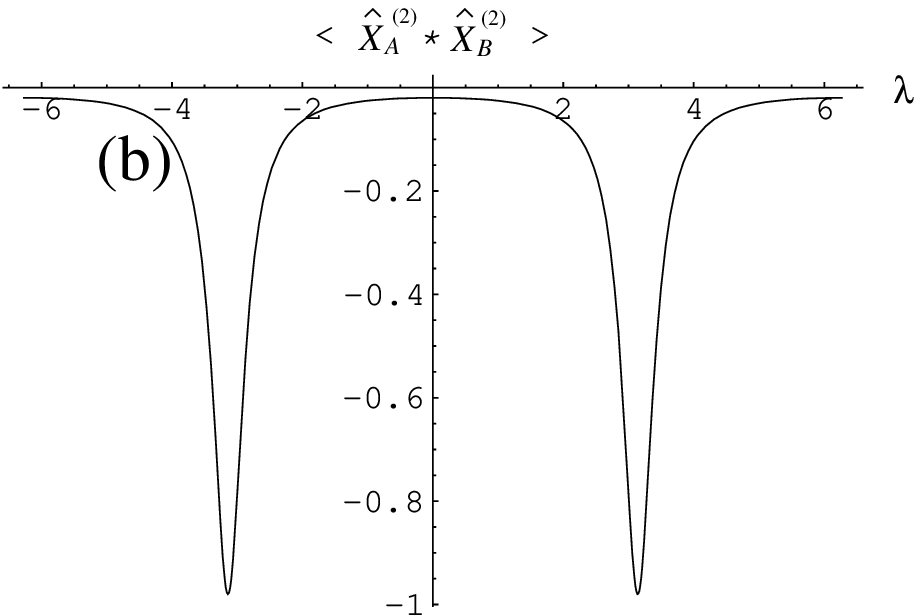}}
\caption{The phase quadrature correlations, $\langle \hat X_{A}^{(2)}
  \hat X_{B}^{(2)}\rangle$, for (a) $\alpha = 1$ and (b) $\alpha =
  0.1$ against $\lambda$.}
\label{quadratue_2}
\end{center}
\end{figure}

Let us now suppose that the initial entangled state shared by Alice
and Bob for a Bell-inequality test is $|\Phi_{+} \rangle_{AB}$, where
$A$ and $B$ denote Alice and Bob's modes, respectively.  If Alice and
Bob perform the local operations, $\hat R(\phi)$ and $\hat R(\theta)$,
on their modes with angles $\phi$ and $\theta$ respectively, the
``rotated'' ECS becomes
\begin{equation}
|\Psi^{R}\rangle_{AB} = N\left\{\cos (\phi-\theta) |\Phi_{+}
\rangle_{AB} + \sin (\phi-\theta) |\Psi_{-}\rangle_{AB} \right\},
\label{rcs}
\end{equation}
where the normalization factor is given by $N = \left\{2\left(1+
\cos[2(\phi - \theta)]e^{-4|\alpha|^2}\right)\right\}^{-1/2}$. The
local operation $\hat R(\varphi)$ changes the normalization factor of
the initial state because of its nonunitary property.  This means the
operation is intrinsically nondeterministic. In this section we ignore
this fact and make the unphysical assumption that Alice and Bob can
apply $\hat R(\varphi)$ deterministically. However, notice that if the
overlap between the two coherent states is negligible, the
normalization factor is not changed by the local operation regardless
of angles $\phi$ and $\theta$. A physically realizable version of
$\hat R(\varphi)$ will be introduced in section III.

First let us examine the correlations between Alice and Bob. The
nonclassical correlations in continuous variable states of light are
described by the electric field amplitude and phase quadratures.  The
amplitude and phase quadratures can be defined respectively as
\noindent
\begin{eqnarray}
\hat X^{(1)}_A = \hat a+\hat a^{\dagger}, \;
\hat X^{(2)}_A = i(\hat a^{\dagger}-\hat a),
\end{eqnarray}
where $\hat a$ and $\hat a^\dagger$ are the field annihilation and
creation operators, and $[\hat X^{(1)}_A,\hat X^{(2)}_A]=2i$. Assuming
real value of $\alpha$, the quadrature correlations for the state
$|\Psi^{R}\rangle_{AB}$ are given by the formulas
\begin{eqnarray}
\langle \hat X_{A}^{(1)} \hat X_{B}^{(1)}\rangle &=& 4\alpha^2
\frac{\cos[2(\phi - \theta)]}{1+ \cos[2(\phi -
    \theta)]e^{-4\alpha^2}}, 
\label{x1x1} \\ 
\langle \hat X_{A}^{(2)} \hat X_{B}^{(2)}\rangle &=& -4\alpha^2
\frac{e^{-4\alpha^2}}{1+ \cos[2(\phi -
    \theta)]e^{-4\alpha^2}}.
\label{x2x2}
\end{eqnarray}
The amplitude quadrature correlation (\ref{x1x1}) for $\alpha=1$ and
$\alpha=0.1$ is depicted in Fig.~\ref{quadratue_1}.  For $\alpha=1$ it
is maximally correlated for $2(\phi - \theta) = 0,\pi,2\pi$ and is
equal to either $-4$ or $4$.  The maximal value increases with
$\alpha$. For $\alpha=0.1$ the correlation is maximal for $2(\phi -
\theta) = \pi$ and is equal to $-1$.  The phase quadrature correlation
(\ref{x2x2}) for $\alpha=1$ and $\alpha=0.1$ is shown in
Fig.~\ref{quadratue_2}. This correlation behaves
  similarly as the amplitude quadrature correlation however, for
  $\alpha=1$ its amplitude is much smaller. This correlation is
  maximal for $2(\phi - \theta) = \pi$ and is equal to $-0.0745$. For
  $\alpha=0.1$ the correlation is maximal for $2(\phi - \theta) = \pi$
  as well and is equal to $-1$. The phase quadrature correlation tends
  to zero if $|\alpha\rangle$ and $|-\alpha\rangle$ tend to orthogonal
  ($\alpha\rightarrow\infty$).

\begin{figure}[h!]
\begin{center}
\scalebox{0.75}{\includegraphics{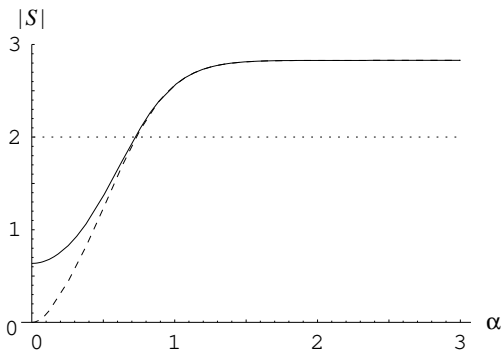}}
\caption{The numerically optimized Bell parameter $|S|$ (solid line)
  and the Bell parameter for $\theta_1=\pi/8$, $\theta_2=3\pi/8$,
  $\phi_1=\pi/4$, $\phi_2=0$ (dashed line).  The Bell parameter $S$
  exceeds the local bound, $2$, for $\alpha\ge 0.723$ and reaches up
  to $2\sqrt{2}$ as $\alpha$ increases.}
\label{S} 
\end{center}
\end{figure}

The amplitude quadrature, in particular, shows high visibility fringes
for $\alpha =1$. However, in order to test a Bell inequality we need
to discretize Alice and Bob's results. After applying the local
operations, $\hat R(\phi)$ and $\hat R(\theta)$, Alice and Bob perform
amplitude homodyne detection on modes $A$ and $B$, respectively.  If
the outcome of Alice's (Bob's) homodyne measurement is larger than 0,
value 1 is assigned to $a$ ($b$).  On the other hand, if Alice's
(Bob's) outcome is smaller than 0, $-1$ is assigned to $a$ ($b$).  The
Bell parameter $S$ is then defined as
\begin{equation}
 S = \langle a_1 b_1\rangle + \langle a_1 b_2\rangle + \langle a_2
 b_1\rangle - \langle a_2 b_2\rangle,
\label{eq:para}
\end{equation}
where the correlation coefficient $\langle a_j b_k\rangle$ corresponds
to the average value of Alice and Bob's joint measurement and the
subscript $j$ ($k$) denotes that angle $\phi_j$ ($\theta_k$) is
applied for the corresponding local operation.  According to any
local-realistic theory, the Bell parameter $S$ should obey the
Bell-CHSH inequality, $|S| \le 2$.  The correlation coefficient
$\langle a_j b_k\rangle$ can be expressed as~\cite{Garcia-Patron2004}
\begingroup \setlength{\arraycolsep}{0pt}
\begin{equation}
\langle a_j b_k\rangle = \int_{-\infty}^{\infty} \!
\mathrm{sign}\left(\eta_{Ar}^j \eta_{Br}^k \right)\, {\cal
  P}\left(\eta_{Ar}^j,\eta_{Br}^k \right) d\eta_{Ar}^j\, d\eta_{Br}^k,
\end{equation}
\endgroup where $\eta_A=\eta_{Ar} + i\eta_{Ai}$ and $\eta_B=\eta_{Br}
+ i\eta_{Bi}$ are the quadrature variables and ${\cal
  P}\left(\eta_{Ar},\eta_{Br} \right)$ is the marginal probability
distribution of the total Wigner function of the state
$|\Psi^{R}\rangle_{AB}$.  The total Wigner function can be calculated
from the characteristic function
\begin{equation}
\chi\left(\zeta_{A},\zeta_{B} \right)= {\rm Tr}\left\{
|\Psi^{R}\rangle\langle\Psi^{R}|_{AB}\,\hat D(\zeta_{A})\otimes
\hat D(\zeta_{B})\right\}
\label{ZZZ1}
\end{equation}
where $\hat D(\zeta)$ is the displacement operator, $\hat
D(\zeta)=\exp [\zeta \hat{a}^{\dagger }-\zeta ^{*}\hat{a}]$, for
bosonic operators $\hat{a}$ and $\hat{a}^{\dag }$.  The Wigner
function is then calculated by taking the Fourier transform of the
characteristic function as
\begin{equation}
\begin{aligned}
&W\left(\eta_A,\eta_B \right)
=\frac{1}{\pi^4}
\int d^2\zeta_A d^2\zeta_B~
 \chi\left(\zeta_{A},\zeta_{B} \right)\times\\
&~~~~~~~~~~~~ \exp[\zeta_A^*\eta_A-\zeta_A\eta_A^*
+\zeta_B^*\eta_B-\zeta_B\eta_B^*].
\label{ZZZ2}
\end{aligned}
\end{equation}
One can calculate the marginal probability distribution ${\cal
  P}(\eta_{Ar}, \eta_{Br})$ using Eqs.~(\ref{bell1}), (\ref{bell2}),
(\ref{rcs}), (\ref{ZZZ1}) and (\ref{ZZZ2}) as
\begin{widetext}
\begin{eqnarray}
{\cal P}(\eta_{Ar}, \eta_{Br}) &=& \int_{-\infty}^{\infty}\, W(\eta_A,
\eta_B) d\eta_{Ai} d\eta_{Bi} 
\nonumber\\
&=&\frac{2}{\pi}N^2 \left\{\cos^2(\phi_j - \theta_k) \left(
e^{-2(\eta_{Ar}-\alpha)^2-2(\eta_{Br}-\alpha)^2} \right.\right.  +
\left.\left. e^{-2(\eta_{Ar}+\alpha)^2-2(\eta_{Br}+\alpha)^2}\right)\right.
\nonumber\\ 
&+& \left. \sin^2(\phi_j - \theta_k) \left(
e^{-2(\eta_{Ar}-\alpha)^2-2(\eta_{Br}+\alpha)^2} \right.\right.  +
\left.\left. e^{-2(\eta_{Ar}+\alpha)^2-2(\eta_{Br}-\alpha)^2}\right)\right.
\nonumber\\
&+& \left. 2 \cos[2(\phi_j - \theta_k)]
e^{-2(\eta_{Ar}^2-\alpha^2)-2(\eta_{Br}^2-\alpha^2) - 8\alpha^2}
\right.  + \left.  \sin[2(\phi_j - \theta_k)]e^{-2\alpha^2}
\left(e^{-2(\eta_{Ar}-\alpha)^2 - 2(\eta_{Br}^2-\alpha^2)}
\right. \right.  
\nonumber\\
&-& \left. \left. e^{-2(\eta_{1r}+\alpha)^2 - 2(\eta_{Br}^2-\alpha^2)}
\right. \right. + \left. \left. e^{-2(\eta_{Ar}^2-\alpha^2) -
  2(\eta_{Br}+\alpha)^2} \right.\right.  -
\left. \left. e^{-2(\eta_{Ar}^2-\alpha^2) - 2(\eta_{Br}-\alpha)^2}
\right) \right\}. 
\label{marginal}
\end{eqnarray}
\end{widetext}
The correlation coefficient evaluated using Eq.~(\ref{marginal}) is
\begin{equation}
\langle a_j b_k\rangle = \frac
{\big(\mathrm{Erfc}[\sqrt{2}\alpha]\big)^2}
{e^{-4|\alpha|^2}+\sec[2(\phi_j-\theta_k)]},
\end{equation}
which is obviously $\alpha$-dependent.  We have numerically found
maximum values of $|S|$ using the method of steepest descent
\cite{nume} and plotted them in Fig.~\ref{S}.  For $\alpha >> 1$ the
figure shows a Bell violation tending to the maximum allowed value of
$2\sqrt{2}$. This is also the regime in which it is valid to treat
$\hat R(\varphi)$ as a unitary. The absolute Bell parameter $|S|$
exceeds the local bound, 2, for $\alpha\geq0.723$.  Angles
$\theta_1=\pi/8$, $\theta_2=3\pi/8$, $\phi_1=\pi/4$, $\phi_2=0$ are
the angles that approximately optimize the violations for
$\alpha\geq0.723$ however, for $\alpha \approx 1$ it is not valid to
treat $\hat R(\varphi)$ as a unitary. To study this region we need to
introduce a physical implementation of $\hat R(\varphi)$. We do this
in the next section.

\section{The local operations for the Bell-inequality tests}

The local operation $\hat R(\varphi)$ required for the Bell inequality
tests studied in the previous section corresponds to a single qubit
rotation for a coherent-state qubit ${\cal A}|\alpha\rangle+{\cal
  B}|-\alpha\rangle$ \cite{JK}.  The $z$-rotation
\begin{equation}
\hat U_z(\varphi)=\left(\begin{array}{cc}
e^{i\varphi} & 0   \\
0 & e^{-i\varphi} 
\end{array}\right)
\end{equation}
for a logical qubit $|\phi\rangle$ can be obtained using the
displacement operator \cite{JK,Ralph}.  The action of the displacement
operator $\hat D(i\epsilon)$, where $\epsilon$ ($\ll \alpha$) is real,
on the qubit $|\phi\rangle$ is approximately the same as the
$z$-rotation of the qubit by $\hat U_z(2\alpha\epsilon)$ when $\alpha
>> 1$.  We can estimate their similarity by calculating the fidelity
\begingroup \setlength{\arraycolsep}{0pt}
\begin{eqnarray}
\label{fidelity-rotation}
&&F=|\langle
\phi|\hat U_z^\dag(2\alpha\epsilon) \hat D(i\epsilon)|\phi\rangle|^2
\nonumber\\ 
&&~=e^{-\epsilon^2}\big\{|{\cal A}|^2+|{\cal
  B}|^2+e^{-2\alpha^2}({\cal AB^*}e^{-2i\alpha\epsilon}+{\cal
  A^*B}e^{2i\alpha\epsilon}) \big\}^2
\nonumber\\ 
&&~\approx \exp[-\epsilon^2]\approx1,
\end{eqnarray}
\endgroup 
where $\alpha\gg1$ was assumed.  The rotation angle $\varphi$ depends
on $\alpha$ and $\epsilon$ as $\varphi=2\alpha\epsilon$.  A small
amount of $\epsilon$ suffices to make one cycle of rotation when
$\alpha$ is relatively large.  The maximum rotation angle is $\pi$
because any angle larger than $\pi$ can be applied to the minus-sign
direction.  In order to make the fidelity to be $F>0.99$ regardless of
the rotation angle, the amplitude should be $\alpha> 15.7$.  It is
well known that the displacement operation $\hat D(i\epsilon)$ can be
effectively performed using a beam splitter with the transmission
coefficient $T$ close to unity and a high-intensity coherent field.

\begin{figure}
\centerline{\scalebox{0.45}{\includegraphics{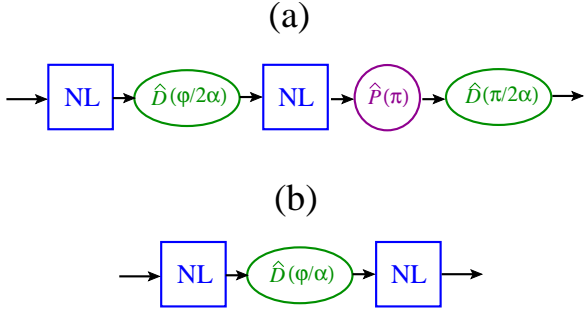}}}
\caption{(a) A schematic of an approximate qubit rotation $\hat
  R(\varphi)$ using Kerr nonlinearities (NL),
  displacement operations ($\hat D$) and a phase shifter ($\hat P$).
  (b) Another example, $\hat V^\prime(\varphi)$, which is an
  approximation of the ideal qubit rotation $\hat R^\prime(\varphi)$.
  See text for details.}
\label{fig:hp}
\end{figure}

To achieve the operation $\hat R(\varphi)$ we need to operate $\hat
U_x(\pi/4)$ and $\hat U_x(-\pi/4)$.  The unitary operation $\hat
U_x(\pi/4)$ can be realized using a Kerr nonlinear interaction
\cite{JK,Yurke}.  The interaction Hamiltonian of a single-mode Kerr
nonlinearity is ${\cal H}_{NL}= \hbar\Omega(a^\dag a)^2$, where
$\Omega$ is the strength of the Kerr nonlinearity.  When the
interaction time $t$ in the medium is $\pi/\Omega$, coherent states
$|\alpha\rangle$ and $|-\alpha\rangle$ evolve to
\begin{eqnarray}
&&\hat U_{NL}|\alpha\rangle
=\frac{e^{-i\pi/4}}{\sqrt{2}}(|\alpha\rangle+i|-\alpha\rangle),\\
&&\hat U_{NL}|-\alpha\rangle
=\frac{e^{-i\pi/4}}{\sqrt{2}}(i|\alpha\rangle+|-\alpha\rangle),
\end{eqnarray}
where $\hat U_{NL}=\exp[i {\cal H}_{NL}t/\hbar]$.  This transformation
corresponds to $\hat U_x(\pi/4)$ up to a global phase shift.  The
other rotation $\hat U_x(-\pi/4)$ can be realized by applying a
$\pi$-phase shifter, $\hat P(\pi)$, which acts
$|\alpha\rangle\leftrightarrow|-\alpha\rangle$, after $\hat
U_x(\pi/4)$ operation.  Since the operation $\hat R(\varphi)$ is
\begin{eqnarray}
\hat R(\varphi) = \hat U_z(\pi) \hat U_x(-\pi/4) \hat U_z(\varphi)\hat
U_x(\pi/4),
\end{eqnarray}
it can be realized using Kerr nonlinearities and linear optics elements
as shown in Fig.~\ref{fig:hp}(a).

It turns out that in order to observe the Bell violation it is
sufficient for Alice and Bob to implement the operation $\hat
V(\varphi,\alpha)$
\begin{equation}
\hat V(\varphi,\alpha) = \hat U_{NL} \hat D\left(
\frac{i\varphi}{\alpha}\right) \hat U_{NL},
\end{equation}
which is depicted in Fig.~\ref{fig:hp}(b).  The operation $\hat
V(\varphi,\alpha)$ is an approximation of the ideal rotation $\hat
R^\prime(\varphi)$
\begin{equation}
\begin{aligned}
&\hat R^\prime(\varphi)|\alpha\rangle = \sin 2\varphi  |\alpha\rangle +
\cos 2\varphi |-\alpha\rangle, \\
&\hat R^\prime(\varphi)|-\alpha\rangle = \cos 2\varphi |\alpha\rangle
 - \sin 2\varphi |-\alpha\rangle,
 \label{e:R_2}
\end{aligned}
\end{equation}
which results in the same Bell inequality violations as with $\hat
R(\varphi)$ in Eq.~(\ref{e:R}). Of course, when $\alpha$ is small, the
``real'' operation $\hat V(\varphi,\alpha)$ is not a good
approximation of $\hat R^\prime(\varphi)$ because the displacement
operator is not a good approximation of $\hat U_z(\varphi)$ in this
limit.  It is straightforward to calculate that \begingroup
\setlength{\arraycolsep}{0cm}
\begin{eqnarray}
&&\hat V(\varphi,\alpha) |\alpha\rangle = \frac{1}{2}
\left\{e^{i\varphi}\Big(|\alpha + \frac{i\varphi}{\alpha}\rangle
+i |-\alpha - \frac{i\varphi}{\alpha}\rangle \Big)  \right. 
\nonumber\\
&&~~~~~~~~~+ \left. i e^{-i\varphi}\Big(|-\alpha + \frac{i\varphi}{\alpha}\rangle
+i |\alpha - \frac{i\varphi}{\alpha}\rangle\Big)
 \right\}, 
\nonumber \\ 
&&\hat V(\varphi,\alpha) |-\alpha\rangle = \frac{1}{2}
\left\{ie^{i\varphi}\Big(|\alpha + \frac{i\varphi}{\alpha}\rangle
+i |-\alpha - \frac{i\varphi}{\alpha}\rangle \Big)  \right. 
\nonumber\\
&&~~~~~~~~~+ \left.  e^{-i\varphi}\Big(|-\alpha +\frac{i\varphi}{\alpha}\rangle
+i |\alpha - \frac{i\varphi}{\alpha}\rangle\Big)
 \right\}.
\label{eq:rotation_unitary}
\end{eqnarray}
\endgroup

If Alice and Bob perform the local operations, $\hat V(\phi,\alpha)$
and $\hat V(\theta,\alpha)$, on their modes of $|\Phi_{+}
\rangle_{AB}$ respectively, the ECS is transformed to
$|\Psi^{V}\rangle_{AB}$ as
\begin{eqnarray}
|\Psi^{V}\rangle_{AB} &=& \frac{N}{2}\left\{
e^{i(\phi-\theta)} (-|\beta_\phi,\gamma_\theta\rangle + i
|\beta_\phi,-\gamma_\theta\rangle \right.  
\nonumber\\ 
&&- \left. i
|-\beta_\phi,\gamma_\theta\rangle - |-\beta_\phi,-\gamma_\theta\rangle) \right.
\nonumber\\ 
&&+ \left. e^{-i(\phi-\theta)}
(-|\gamma_\phi,\beta_\theta\rangle - i |\gamma_\phi,-\beta_\theta\rangle \right.
\nonumber\\ 
&&+ \left.i |-\gamma_\phi,\beta_\theta\rangle -
|-\gamma_\phi,-\beta_\theta\rangle)\right\},
\label{new_rcs}
\end{eqnarray}
where $\beta_{\phi,\theta} = \alpha + i\varphi_{\phi,\theta}/\alpha$,
$\gamma_{\phi,\theta} = \alpha -i\varphi_{\phi,\theta}/\alpha$.  The
Bell parameter $S$ in Eq.~(\ref{eq:para}) can then be obtained using
the Wigner representation of Eq.~(\ref{new_rcs}), as described in the
previous section.

\begin{figure}
\centerline{\scalebox{0.75}{\includegraphics{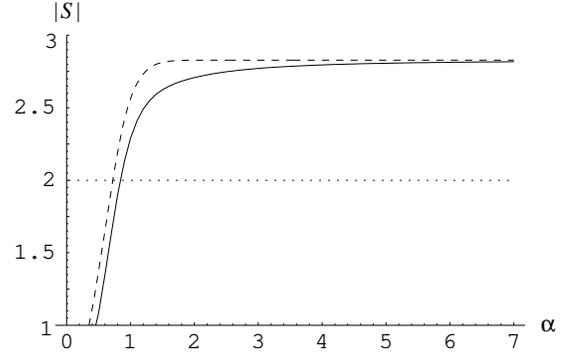}}}
\caption{ The numerically optimized Bell parameter $|S|$ for an ECS
  with amplitude $\alpha$ using the real local operation (solid curve)
  and the ideal local operation (dashed curve).  See text for
  details.}
\label{fig:double}
\end{figure}

The explicit expressions of the Wigner function of state
$|\Psi^{V}\rangle_{AB}$ and its Bell parameter $S$ are inappropriate
to present here since they are too lengthy.  In Fig.~\ref{fig:double},
we have plotted the absolute Bell parameter $|S|$ of the ECS maximized
using the method of steepest descent \cite{nume} (solid curve) and
compare it with the case using the ``ideal'' rotation (dashed curve).
The violations reach up to Cirel'son's bound $2\sqrt{2}$, as $\alpha$
grows.  Remarkably, the Bell violations of the ECS using the ``real''
operation $\hat V(\varphi_a,\alpha)$ does not require very large
values of $\alpha$.  The Bell inequality is violated for $\alpha\geq
0.842$ using $\hat V(\varphi_a,\alpha)$ while it was $\alpha\geq
0.723$ when the unphysical idealized local operation $\hat R(\varphi)$
was applied.  In the case of an ECS with $\alpha =1$ the maximum
violation is $S\approx 2.29$ at $\theta_1\approx-0.066$,
$\theta_2\approx0.066$ $\phi_1\approx0.236$, $\phi_2\approx-0.236$.

\section{Conclusion}

In this paper we have studied the Bell-CHSH inequality with ECSs,
local non-linear operations and homodyne measurements.  An ECS with a
large amplitude is a state which contains quantum correlations
between macroscopically distinguishable states.  Optical states are
considered macroscopically distinguishable if
they can be distinguished by homodyne detection. We have shown that
the Bell-CHSH inequality can be violated with ECSs using homodyne
measurements up to Tsirelson's bound $2\sqrt{2}$. The bound is
approached when $\alpha \gg 1$. Surprisingly, violation of local
reality with respect to homodyne measurements persists down to
$\alpha\geq 0.842$.

Given the importance of entanglement from both a fundamental
perspective and that of applications such as quantum computing, it
would be of considerable interest to test these ideas
experimentally.
In order to generate an ECS with $\alpha =1$ a single mode
SCS with amplitude
$\alpha=\sqrt{2}\approx 1.414$ is required. For this ECS a
maximum value of the Bell parameter is $\approx 2.29$ which is
significantly larger than the classical limit.
  Production of
ECSs of this size are within reach of current technology
 \cite{g-1,g-15,MP,g-2}.
It is known that a SCS with a small amplitude
is very well approximated by a squeezed single photon \cite{g-1}.
A SCS with amplitude $\sqrt{2}$ and fidelity
$\approx 0.97$ may be produced by squeezing a single photon
with 4.8dB squeezing, which is experimentally feasible.

The strength of the Kerr
non-linearity required for the local operations remains challenging
however, efforts are being made to obtain nonlinear effects of
sufficient strength using electromagnetically induced transparency
\cite{Hau,MP,Pet}. It should be noted that ECSs with small amplitudes,
which we are interested in for experimental realization,
are relatively less sensitive to noise during the nonlinear
interactions.

The experimental realization
of Bell violations with large amplitudes, $\alpha\gg1$,
would be even more interesting since in this limit,
ECSs can be considered to be truly ``macroscopic" entanglement.
As shown in Fig.~5, Bell inequality violations close to
the Cirel'son's bound occur for $\alpha\gg1$.
There are some technical difficulties in approaching this regime
experimentally.
Firstly, it is known that 
the generation of an ECS of an amplitude $\alpha \gg 1$
is experimentally more demanding.
However, some recent theoretical proposals are expected to
be experimentally implemented in foreseeable future
to generate ECSs with large amplitudes.
For example, one may use the SCS amplification scheme \cite{g-1},
which uses beam splitters, ancillary coherent states and photodetectors,
to distill large SCSs out of small ones.
It was shown that a SCS of $\alpha\approx2.5$, which means an ECS of
$\alpha\approx1.8$ may be realized using experimentally available resources
with a high fidelity \cite{g-1}.
Secondly, in the case of a large $\alpha$, the local operations
will be harder to be performed.
When amplitudes of ECSs are larger, they suffer more rapid 
destruction of quantum coherence in the nonlinear media used
for the local operations. 
Methods to efficiently perform the local operations for our
Bell inequality tests deserve further investigations.

\acknowledgments This work was partially supported by a~MEN Grant
No.~1 PO3B 137 30, N202 021 32/0700, the DTO-funded U.S. Army Research
Office Contract No. W911NF-05-0397, the Australian Research Council and 
Queensland State Government.


\begin{thebibliography}{99}
\bibitem{error} The error rate for discriminating between two coherent
  states, $|\alpha\rangle$ and $|-\alpha\rangle$, can be calculated as
  $1-\int_0^\infty dx \exp[-(x-\sqrt{2}\alpha)^2]/\sqrt{\pi}$.
\end{thebibliography}
\end{document}